\documentclass[aps,prc,twocolumn]{revtex4-1}

\usepackage{bm,graphicx,textcomp,amssymb,amsmath,dcolumn,color}

\newcommand{\f}[1]{\mbox{\boldmath$#1$}}

\newcommand{\bea}{\begin{eqnarray}}
\newcommand{\ea}{\end{eqnarray}}
\newcommand{\eea}{\end{eqnarray}}

\begin{document}

\title{Comment on ``Enhanced deuterium-tritium fusion cross sections 
in the presence of strong electromagnetic fields''}

\author{Friedemann Queisser and Ralf Sch\"utzhold}

 \affiliation{Helmholtz-Zentrum Dresden-Rossendorf, 
 Bautzner Landstra{\ss}e 400, 01328 Dresden, Germany,}
 
 \affiliation{Institut f\"ur Theoretische Physik, 
 Technische Universit\"at Dresden, 01062 Dresden, Germany.} 

\date{\today}

\begin{abstract}
In their article [Phys.\ Rev.\ C {\bf 100}, 064610 (2019)], Lv, Duan, and Liu 
study the enhancement of deuterium-tritium fusion reactions by the 
electromagnetic field of an x-ray free-electron laser (XFEL).
While we support the general idea (which was put forward earlier in our 
rapid communication [Phys.\ Rev.\ C {\bf 100}, 041601(R) (2019)]),
we find that the time-averaged potential approximation used by Lv, Duan, and Liu 
is not justified in this regime and does not take into account important effects.
Due to those effects, the enhancement mechanism 
may actually be more efficient than predicted by Lv, Duan, and Liu.  
\end{abstract}

\maketitle

In their article \cite{lv+duan+liu}, Lv, Duan, and Liu 
study the enhancement 
of tunneling in deuterium-tritium fusion reactions induced by the 
electromagnetic field of an x-ray free-electron laser (XFEL).
While we fully agree that this is an interesting and potentially important 
subject worth investigating from various angles, 
we believe that a few remarks are in order.
First, the main idea of assisting deuterium-tritium (and other) fusion reactions 
by the strong electromagnetic field of a XFEL has already been put forward 
earlier in our rapid communication \cite{our}. 
Second, and more important, the approximation of tunneling in a time-averaged 
potential (after representing the vector potential $\f{A}(t)$
of the XFEL by a time-dependent spatial translation) 
used in \cite{lv+duan+liu} is not justified in the regime under consideration. 

While our understanding of tunneling -- especially in time-dependent 
scenarios -- is still far from complete, there are a number of known results.
Let us start with the limiting cases.
If the external time-dependence (e.g., of the potential barrier $V$) is very slow, 
i.e., much slower than all relevant time scales for tunneling, we may use the 
quasi-static approximation by neglecting the external temporal dependence during 
the tunneling process. 
This regime is discussed in the paragraph {\em Deformation of potential} 
in \cite{our}. 
In the other limiting case, when the external oscillations of $V(t,\f{r})$
are not too violent and very fast, i.e., much faster than all other relevant 
frequency or energy scales, 
one may consider an effectively time-averaged potential 
$V(t,\f{r})\to\bar{V}_{\rm eff}(\f{r})$, 
which is the approximation used in \cite{lv+duan+liu}. 

However, in between these two limiting cases, there is plenty of room for 
rich physics and many fascinating phenomena.
One of them is the Franz-Keldysh effect \cite{Franz,Keldysh} describing 
changes of the tunneling rates by oscillating external fields. 
Employing Floquet analysis, the first Floquet side-bands effectively behave 
as waves with larger or smaller energy ${\cal E}\to{\cal E}\pm\hbar\omega$ 
and thus lead to frequency dependent tunneling rates, cf.~\cite{our}. 
As a consequence, the relevant frequency scale for the Franz-Keldysh effect 
is set by characteristic energy scales of potential barrier and the resulting 
behavior of the tunneling wave functions. 

Another important time scale is the B\"uttiker-Landauer traversal time $\tau$
which is precisely motivated by the question: ``When does a time-dependence 
have an effect on the tunneling probability?'', see \cite{Buttiker}. 
Within the instanton picture (i.e., going to imaginary time), this time 
scale $\tau$ is set by the period of oscillation in the potential barrier 
turned upside-down. 
Hence, the associated frequency scale~$\sim1/\tau$ may be very different 
from the characteristic frequency scales $\omega$ of the Franz-Keldysh effect. 

In view of all these effects, a threshold frequency of 1~keV above which 
the time-averaged potential approximation $V(t,\f{r})\to\bar{V}_{\rm eff}(\f{r})$
is supposed to apply, as assumed in \cite{lv+duan+liu}, appears far to low.
The comparably long period of several femto-seconds (i.e., in the optical regime) 
referred to as ``collision time'' in \cite{lv+duan+liu} is not the only relevant 
time scale. 
Neglecting small corrections due to the finite size of the nuclei, 
the B\"uttiker-Landauer traversal time $\tau$ for tunneling in 
deuterium-tritium fusion reads 
\bea 
\tau
=
\hbar\,\frac{\pi}{4}\sqrt{\frac{2\mu c^2}{{\cal E}^3}}\,\alpha_{\rm QED}
\,,
\ea
where $\cal E$ is the initial kinetic energy, $\mu$ the reduced mass, 
and $\alpha_{\rm QED}\approx1/137$ the fine-structure constant \cite{our}.
For an energy of ${\cal E}=4~\rm keV$, for example, 
the inverse B\"uttiker-Landauer traversal time $1/\tau$ is also around 1~keV 
which means that XFEL frequencies in the keV regime 
are just in the right range to probe these interesting 
(and strongly frequency dependent) effects -- which are not captured by 
the time-averaged potential approximation.  
For higher energies $\cal E$, the B\"uttiker-Landauer traversal time $\tau$
would be even smaller. 
Thus, keeping the XFEL frequency fixed at 1~keV and increasing the energy $\cal E$, 
one would move towards the regime of applicability of the quasi-static approximation 
and even further away from the limit where the time-averaged potential yields a good 
approximation.  

Furthermore, the nuclear energy barrier height of 0.37~MeV and well depth 
between 30 and 40~MeV facilitate a huge number of phase oscillations of 
the wave function during one XFEL period (for a frequency of 1~keV), 
which -- together with the steep slope of the potential in between -- 
casts further doubts on the applicability of the time-averaged potential 
approximation. 
Note that the process of tunneling is non-perturbative in terms of the 
coupling to the potential $V$, which implies that special care is required 
for approximations involving the potential barrier. 
Thus, it can be advantageous to represent 
the XFEL field by a vector potential $\f{A}(t)$ instead, since this representation 
facilitates a perturbative treatment (within the tunneling exponent), 
as long as the XFEL field is not too strong, see \cite{our}. 

Another point of concern is the nuclear fusion time scale itself,
i.e., the time it takes the deuterium and tritium nuclei 
to interact and to actually fuse, provided that they are close enough 
(i.e., after tunneling through the potential barrier $V$). 
By using the time-averaged potential approximation
$V(t,\f{r})\to\bar{V}_{\rm eff}(\f{r})$, one is implicitly 
assuming that these nuclear fusion scales are also much slower 
than the XFEL oscillation period. 

Apart from the issue of the threshold frequency discussed above, 
the amplitude of the oscillation can also lead to problems.
If, as assumed in \cite{lv+duan+liu}, this quiver amplitude 
(referred to as $r_{\rm e}$ in \cite{lv+duan+liu})
becomes larger than the spatial extend of the nuclei 
(referred to as $r_{\rm n}$ in \cite{lv+duan+liu}) 
the question of whether the wave function is inside or outside 
the potential 
well becomes problematic (which is of course related to the 
points above). 

In fact, as shown in \cite{our}, the phenomena neglected by the 
time-averaged potential approximation 
could actually increase the tunneling probability 
more efficiently and already at lower field strengths \cite{footnote} 
than the mechanism considered in \cite{lv+duan+liu}.  

%\bigskip

In summary, while we do not question the main idea or the validity of the 
model described in Sec.~II of \cite{lv+duan+liu} 
(which are basically the same as in \cite{our}), we would like to point out that 
the time-averaged potential approximation used in \cite{lv+duan+liu} 
is not justified for XFEL frequencies in the keV regime under consideration. 
However, this may actually be good news -- as the enhancement of fusion could 
be more efficient \cite{footnote} 
than expected from the time-averaged potential approximation. 

\acknowledgments 

The authors acknowledge 
fruitful discussions with C.~Kohlf\"urst, R.~Sauerbrey
and other colleagues from the HZDR as well as 
financial support by the German Research Foundation DFG 
(grants 278162697 -- SFB 1242, 398912239).


\begin{thebibliography}{99}
\bibitem{footnote}
For an XFEL frequency of $\omega=1~\rm keV$ (casting aside the problem 
of the applicability of the time-averaged potential approximation), 
the required intensity $I$ one can read off Fig.~1 in \cite{lv+duan+liu} 
is larger than $10^{22}~\rm W/cm^2$.
%
This corresponds to an electric field strength of more than 
$10^{14}~\rm V/m$ which is comparable to the nuclear Coulomb field 
at the turning point (minimum distance) of around 1.4~pm for an 
energy of ${\cal E}=1~\rm keV$.
%
As shown in \cite{our}, modifications of the tunneling process 
(e.g., dynamical assistance) should be taken into account at those 
field strengths. 
%
Note that these modifications display a non-trivial dependence on 
$\omega$ and $I$ etc. 
%
By contrast, in the time-averaged potential approximation, 
$\omega$ and $I$ only enter via the quiver amplitude 
(referred to as $r_{\rm e}$ in \cite{lv+duan+liu}) which scales
as $r_{\rm e}\propto\sqrt{I}/\omega^2$.
%
Thus, for $\omega=10~\rm keV$, the required intensity would be 
above $10^{26}~\rm W/cm^2$ corresponding to an electric field 
strength of more than $10^{16}~\rm V/m$.
%
This exceeds the nuclear Coulomb field mentioned above and 
significantly modifies the tunneling process -- unless very high 
energies $\cal E$ are considered. 
%
(But at such high energies $\cal E$, the problem of the applicability 
of the time-averaged potential approximation becomes even more 
pronounced, as explained above.)
\end{thebibliography}
\end{document}